\documentclass[sigconf]{acmart}
\AtBeginDocument{%
  \providecommand\BibTeX{{%
    \normalfont B\kern-0.5em{\scshape i\kern-0.25em b}\kern-0.8em\TeX}}}

\usepackage{graphicx}
\usepackage{subcaption}
\usepackage{multirow}
\usepackage{booktabs}
\usepackage{tabularx}
\usepackage{bm}
\usepackage{url}
\usepackage{tablefootnote}
\usepackage{threeparttable}

\usepackage{hyperref}
\hypersetup{colorlinks,allcolors=black}

\copyrightyear{2020} 
\acmYear{2020} 
\setcopyright{rightsretained} 
\acmConference[ICMI '20]{Proceedings of the 2020 International Conference on Multimodal Interaction}{October 25--29, 2020}{Virtual event, Netherlands}
\acmBooktitle{Proceedings of the 2020 International Conference on Multimodal Interaction (ICMI '20), October 25--29, 2020, Virtual event, Netherlands}

\begin{document}
\fancyhead{}

\title{Dyadic Speech-based Affect Recognition using \\DAMI-P2C Parent-child Multimodal Interaction Dataset}

\author{Huili Chen}
\affiliation{%
  \streetaddress{MIT Media Lab}
  \city{MIT Media Lab}}
\email{hchen25@media.mit.edu}

\author{Yue Zhang}
\affiliation{%
  \city{MIT Media Lab}}
\email{yuefw@mit.edu}

\author{Felix Weninger}
\affiliation{%
  \city{Nuance Communications}}
\email{felix@weninger.de}

\author{Rosalind Picard}
\affiliation{%
  \city{MIT Media Lab}}
\email{picard@media.mit.edu}

\author{Cynthia Breazeal}
\affiliation{%
  \city{MIT Media Lab}}
\email{cynthiab@media.mit.edu}

\author{Hae Won Park}
\affiliation{%
  \city{MIT Media Lab}}
\email{haewon@media.mit.edu}



\begin{abstract}
Automatic speech-based affect recognition of individuals in dyadic conversation is a challenging task, in part because of its heavy reliance on manual pre-processing. Traditional approaches frequently require hand-crafted speech features and segmentation of speaker turns. In this work, we design end-to-end deep learning methods to recognize each person's affective expression in an audio stream with two speakers, automatically discovering features and time regions relevant to the target speaker's affect. We integrate a local attention mechanism into the end-to-end architecture and compare the performance of three attention implementations -- one mean pooling and two weighted pooling methods. Our results show that the proposed weighted-pooling attention solutions are able to learn to focus on the regions containing target speaker's affective information and successfully extract the individual's valence and arousal intensity. Here we introduce and use a ``dyadic affect in multimodal interaction - parent to child'' (DAMI-P2C) dataset collected in a study of 34 families, where a parent and a child (3-7 years old) engage in reading storybooks together. In contrast to existing public datasets for affect recognition, each instance for both speakers in the DAMI-P2C dataset is annotated for the perceived affect by three labelers. To encourage more research on the challenging task of multi-speaker affect sensing, we make the annotated DAMI-P2C dataset publicly available\footnote{The database is available on \url{https://forms.gle/FXJAPRUgkLwbW8NN9}.}, including acoustic features of the dyads' raw audios, affect annotations, and a diverse set of developmental, social, and demographic profiles of each dyad. 
\end{abstract}

\ccsdesc[500]{Affective computing and interaction}
\ccsdesc[500]{Speech behaviours in social interaction}
\ccsdesc[500]{Multimodal datasets and validation}

\keywords{dyadic affect dataset, speech affect recognition, parent-child interaction, end-to-end learning}


\maketitle

\section{Introduction}
\subsection{Learning technologies for parent-child dyadic interactions}
High-quality, social, responsive, and facilitative interactions between parents and their children are crucial for young children's language development~\cite{Roseberry2014-SMS,Tamis2001-MRA}. 
A major problem faced by many children, particularly those from low socioeconomic status (SES) families, is limited exposure to rich language during adult-child conversations at home. For instance, studies have reported that in low-SES families, parent-to-child conversations tend to be less frequent and of shorter duration~\cite{Hart1995-MDI,Rowe2008-CDS}, and have fewer open-ended questions and discussions (e.g., parents negotiating with their children)~\cite{Hart1995-MDI,Rowe2008-CDS,Hoff2003-TSO,Lawrence1996-PST} compared higher-SES families.
Such parental “participation gap” is gaining significant attention, as the active participation of parents (or other adults children interact with daily) is crucial for children's educational success~\cite{Neuman2012-WAO,Neuman2015-GOC}. It should be noted that the interest level of parents in their children's education does not vary across different SES groups~\cite{Neuman2012wordsapart}; however, the lack of access to parental education and guidance contributes to the participation gap. As noted by Hoover et al.~\cite{Hoover1995-PII}, a well guided parental involvement in children’s education has a significant impact on children's cognitive development and literacy skills.

Therefore, there is urgency in figuring out how to enrich social and conversational interactions between parents and children, particularly those from lower-SES households. Despite the increasing number of educational apps designed to support parental participation~\cite{Mcnab2013-DTI,Takeuchi2011-TNC}, few artificial intelligence (AI) technologies have been designed to support multimodal and reciprocal adult-child interaction, for example, to proactively facilitate dialogic storytelling between two interlocutors in the here and now -- where parent and child not only read but actively converse about the story, asking and answering questions, commenting on the narrative, and so on~\cite{Chang2011-TSR}. 

To achieve this objective, AI-enabled learning technologies (e.g., robot learning companions~\cite{CHEN2020TLC,Spaulding2018-ASRS}) need to have the ability to accurately recognize affective expressions of individuals in the parent-child dyads. Children's affective states have been shown to be crucial for their learning during their interaction with learning technologies~\cite{Chen2020-IOI}, and affect-aware learning technologies promote children's learning more effectively than affect-blind ones~\cite{Gordon2016APS-APO,Park2019-AMF}. Such affect sensing ability offers opportunities for learning technologies to provide real-time personalized interactions and interventions that optimize the learning experience for each family.  

\subsection{State-of-the-art affect sensing models}
Current speech affect sensing models focus primarily on single-person affect detection and assume that an audio recording containing multiple speakers' utterances is manually partitioned into homogeneous segments according to the speaker's identity.
Thus, speaker diarization has been an inevitable preprocessing step for these models, and as a result, modeling affect of individuals in dyads has been treated as a two-step problem. 
As far as we know, there is currently no approach that offers speaker diarization and affect recognition in a single model. 

The traditional feature designing approach for speech-based affect recognition also creates additional barriers and requires much manual effort.
After a large number of acoustic descriptors, such as pitch and energy, are extracted on a frame level from raw speech signals, feature designing techniques, e.g., statistical functionals, are applied to obtain a subset of most salient features for classification or regression with machine learning algorithms~\cite{Eyben2013-RDI}. The feature attributes for model training are frequently selected through extensive and carefully prepared experiments and their effectiveness heavily relies on the implemented pattern recognition model~\cite{Eyben2013-RDI}. Typically, a researcher invests a lot of manual effort in these steps, with a common belief being that the more effort, the greater the generalizability of the results.

As an increasingly popular alternative approach that can reduce manual effort, end-to-end deep learning automatically crafts a feature set and explores the most salient representations related to the task of interest by jointly training the representation learning and pattern recognition processes~\cite{Trigeorgis2016-AFE,Tzirakis2017-ETE,Li2018-AAP,Sarma2018-EIF,Mirsamadi2017-ASE}. For example, Li et al.~\cite{Li2018-AAP} have applied convolutional neural networks (CNNs) combined with attention pooling directly to the spectrograms extracted from speech utterances to jointly model temporal and frequency domain information and learn a final emotional representation. Instead of using CNNs, Mirsamadi et. al~\cite{Mirsamadi2017-ASE} have designed a deep recurrent neural network (RNN) system that learns both emotionally relevant features and their temporal aggregation from either raw audio signals or low-level descriptors of speech.

To our knowledge, the current end-to-end approaches are, however, designed for single-person affect recognition, and have not been applied to multi-speaker affect recognition. In a multi-speaker scenario, it is important to learn which time regions hold important information for affect prediction of a specific speaker.   
For instance, in an audio where a parent is talking with a child, an affect recognition model for the child's arousal level needs to intelligently identify time regions when the child is speaking and the segments within those time regions most indicative of her arousal level. This means that an end-to-end model needs to have an attention mechanism that can learn the varying importance of the representations at different times as well as focus on the target speaker instead of simply averaging them across an entire audio that may contain utterances from both speakers.
.

\subsection{Our approach and contributions}
In this work, we integrate an attention mechanism into an end-to-end framework as an implicit speaker diarizer with the aim to automatically focus on the target speaker in dyads when modeling that speaker's affect. 

Since its inception, the attention mechanism has been widely used in various speech-related tasks including speech recognition and natural language processing. 
Additionally, more recent state-of-the-art work on emotion recognition started to implement attention mechanisms into end-to-end learning models in various ways and have increased the predictive power of the models~\cite{Mirsamadi2017-ASE,Huang2017-DCR,Zhang2019-AAE,Meng2019-SER}. For example, Mirsamadi~\cite{Mirsamadi2017-ASE} has augmented deep neural networks (DNNs) with a temporal attention mechanism after RNNs to extract the most emotionally-salient parts of speech signals for emotion prediction. Another prior work added attention layers to 
CNNs for emotion recognition~\cite{Li2018-AAP}. 

Here, we apply weighted pooling with local attention to study a novel problem in the speech affect field -- dyadic affect recognition. To the best of our knowledge, this is the first work that is applying attention mechanisms in DNNs to model dyadic affect predictions. 
In fact, there has been very little research effort in recognizing the affect of both participants in a dyadic interaction. One of the main reasons for the lack of research in this field is that there is no existing dyad dataset with proper affect annotations.  Even the most popular dyadic datasets such as 
IEMOCAP~\cite{Busso2008-IIE} does not provide the labels for both speakers' emotions. 
To help advance research in this area, we present a new dataset, ``dyadic affect in multimodal interaction - parent to child'' (DAMI-P2C), which captures the affect (arousal and valence) of both parent and child in natural story-reading dyads. This is the first dataset that provides annotations for both speakers' affective expressions for each instance.

In summary, the main contributions of this work are: (1) A rich dataset of parent-child dyads with affect annotations and other social and developmental profiles of the dyads, (2) The design and development of an end-to-end learning framework augmented with an attention mechanism for multi-speaker affect recognition, and (3) A comprehensive analysis with experimental results to provide a competitive affect recognition baseline for future multi-speaker affect recognition research. 

\section{Dyad Interaction Data collection}\label{data collection}
\subsection{Collection protocol}
\begin{figure}[t]
\centering

    \includegraphics[width=0.5\linewidth]{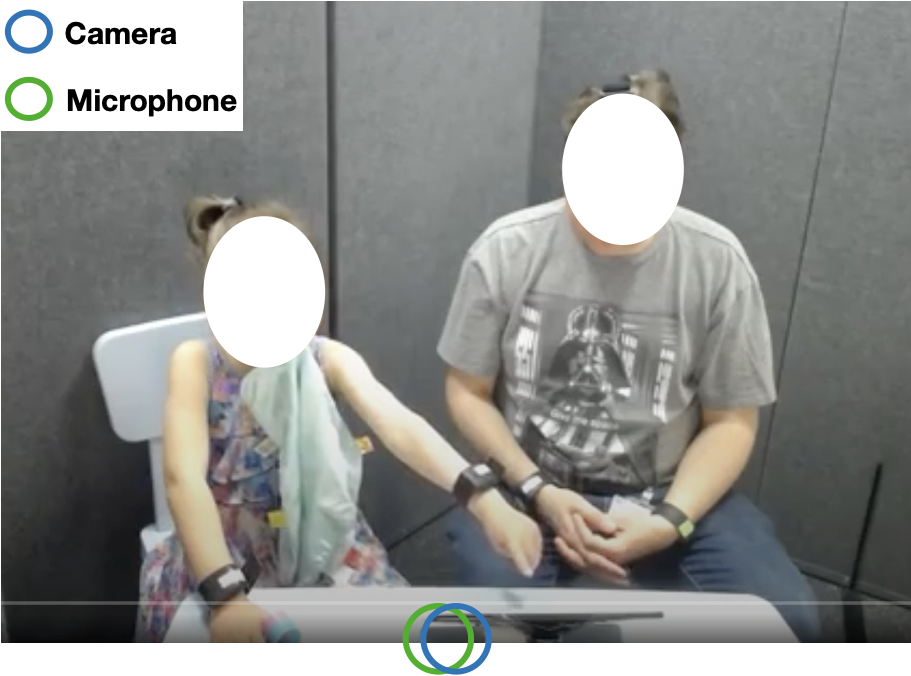}
    \vspace*{-\baselineskip}
    \caption{An example scene of a parent-child storytime interaction in the lab setting. Annotators viewed audio-visual recordings similar to this one, except showing the faces, when coding parent’s and child’s affect attributes. }
    \label{fig:dyadic-interaction}
   \vspace*{-\baselineskip}
\end{figure}

We invited families with children between the ages of 3-7 years to come to our lab space for a two-session data collection. The collection protocol consisted of two 45-minute in-lab sessions where a parent and their child read stories together for around 20 minutes, and the parent filled out surveys for the remaining 25 minutes. During the in-lab sessions, the parent and child sat next to each other as shown in Figure~\ref{fig:dyadic-interaction}. Families that completed both sessions were given \$75 as their compensation. 

A digitized version of our storybook corpus on a touchscreen tablet was used for the sessions. 
The storybook corpus consists of 30 storybooks recommended by early childhood education experts and teachers. Each story lasts from 3 to 15 minutes. Stories shorter than five minutes were categorized as short stories and the rest as long stories. During the story reading session, the parent and child could select any books they wanted to read from the corpus. They were encouraged to read stories together in a way that they would normally do at home, having conversations around the stories and making the activity fun and interactive. 

Thirty-four families were recruited in our data collection from the Greater Boston area with their full consent (Table.~\ref{tab:1:demographic-information}). The dyad was always a pair of one parent and one child. Two parents did not report their children's age, and four parents did not report their own. Three families withdrew from the data collection after the first session for reasons not related to the protocol. In sum, DAMI-P2C dataset consists of two sessions of 31 families and one session of three families.

\begin{table}
\caption{Gender identity, age range and language proficiency of the participant families in the dyadic dataset. English language learner is denoted as ELL. Native/bilingual speaker is denoted as N/B. The average ages for parents and children are $39.70\pm5.47$ and $5.49\pm1.37$, respectively.  }
\label{tab:1:demographic-information}
 \vspace{-4mm}
 \begin{small}
 \begin{tabular}{lrrrr} 
  \toprule
        & parent (age) & child (age) \\ 
  \midrule
  Female &   25 ($38.78\pm4.70$) &  13 ($5.20\pm1.96$)    \\
  Male & 9 ($42.25\pm6.90$) & 21 ($5.65\pm0.96$) \\
  \midrule
  N/B & 28 ($39.54\pm5.15$)  &  29 ($5.58\pm1.46$)   \\
  ELL & 6 ($40.33\pm7.15$) & 5 ($5.00\pm0.64$) \\
  \bottomrule
 \end{tabular}
 \end{small}
  \vspace{-4mm}
\end{table}

\subsection{Measurements}\label{data-measurements}
During the in-lab parent-child co-reading sessions, audio-visual recordings were captured using one microphone and four cameras installed in the story reading station (Figure~\ref{fig:dyadic-interaction}). Four cameras were used to capture different angles of the dyadic interaction (i.e., frontal view, birds-eye view, parent-centered view, and child-centered view). The audio recordings were sent to a professional transcription service to obtain textual annotation of recorded speech. 

In addition, we collected participant families' demographic information, socio-economic status, home literacy environment, parenting style~\cite{Kamphaus2006-PRQ}, a parental theory of mind assessment~\cite{Rice2015-SMC}, and a child's temperament and behavior questionnaire~\cite{Putnam2006-DOS} completed by the parent.~\footnote{The DAMI-P2C dataset is being released in phases, with the first phase having the acoustic features of raw audios, affect annotations, and the developmental, social and demographic profiles of each dyad. We plan to add in the raw audio-visual recordings with faces unobstructed for facial affect analysis in later phases.} The three families who only completed the first in-lab session did not fill out the surveys.

\subsection{Data annotation}

We recruited three trained annotators with a psychology or education background to annotate the audio-visual recordings of the families' co-reading interactions. The affect of both parent and child were annotated separately in terms of valence and arousal.
While watching the recordings, the coders gave ratings every five non-overlapping seconds on a five-point ordinal scale [-2,2], with two corresponding to cases when the target person showed clear signs of high arousal/positive valence and -2 when the person showed clear signs of low arousal/negative valence.~\footnote{Coding manuals for valence/arousal of both parent and child can be accessed via the link (\url{https://drive.google.com/file/d/1wSedbtIZcik8f21AvHU4tKSMOPFU1b5z/view?usp=sharing}).}
A five second window was selected as the fragment interval of target audio-visual recordings for the annotation to produce continuous quality scales, a threshold consistent with prior work on affect detection ~\cite{Rudovic2019-PEE}. When annotating the recordings, annotators were instructed to judge whether a given fragment contained the story-related dyadic interaction and filter out those that did not. For those five-second fragments in which speakers paused and only made utterances in a small time interval, i.e., 1 second, we kept them as valid data instances, since a short period of silence frequently happens naturally in human conversations. In total, 16,593 five-second fragments have been annotated with $488.03\pm123.25$ fragments from each family on average, shown in Table~\ref{des-stats-parent-child-dyads}. 

The agreement of the three annotators was measured using the intra-class correlation (ICC)~\cite{Shrout1979-ICU} type (3,1) for average fixed raters. The ICC is commonly used in behavioral sciences to assess the annotators' agreement and score ranges from 0-100\%. 
The average ICC and its confidence interval among the annotators for each of our labels are presented in Table~\ref{correlation-between-affect-attributes}. According to the commonly-cited cutoffs for qualitative inter-rater reliability (IRR) based on ICC values~\cite{Hallgren2012-ComputingIR,Cicchetti1994-GCR}, IRR is good if $.60 \leq ICC < .75$, and is excellent if $.75 \leq ICC \leq 1.0$. Given this evaluation criteria, the annotations for the four affect attributes all achieve either good or excellent qualities. After recordings were coded independently by the annotators, we took the average of a scale's ratings from the three annotators for each recording fragment as its final score. The ratings were standardized to the zero mean and unit standard deviation for each annotator in order to eliminate individual judgement biases of the three annotators in the use of the rating scale. Both the average standardized ratings and the three independent raw ratings are available in the DAMI-P2C dataset. 

\begin{figure}[t]
\centering
\begin{subfigure}[h]{0.42\linewidth}
\includegraphics[width=1.00\linewidth]{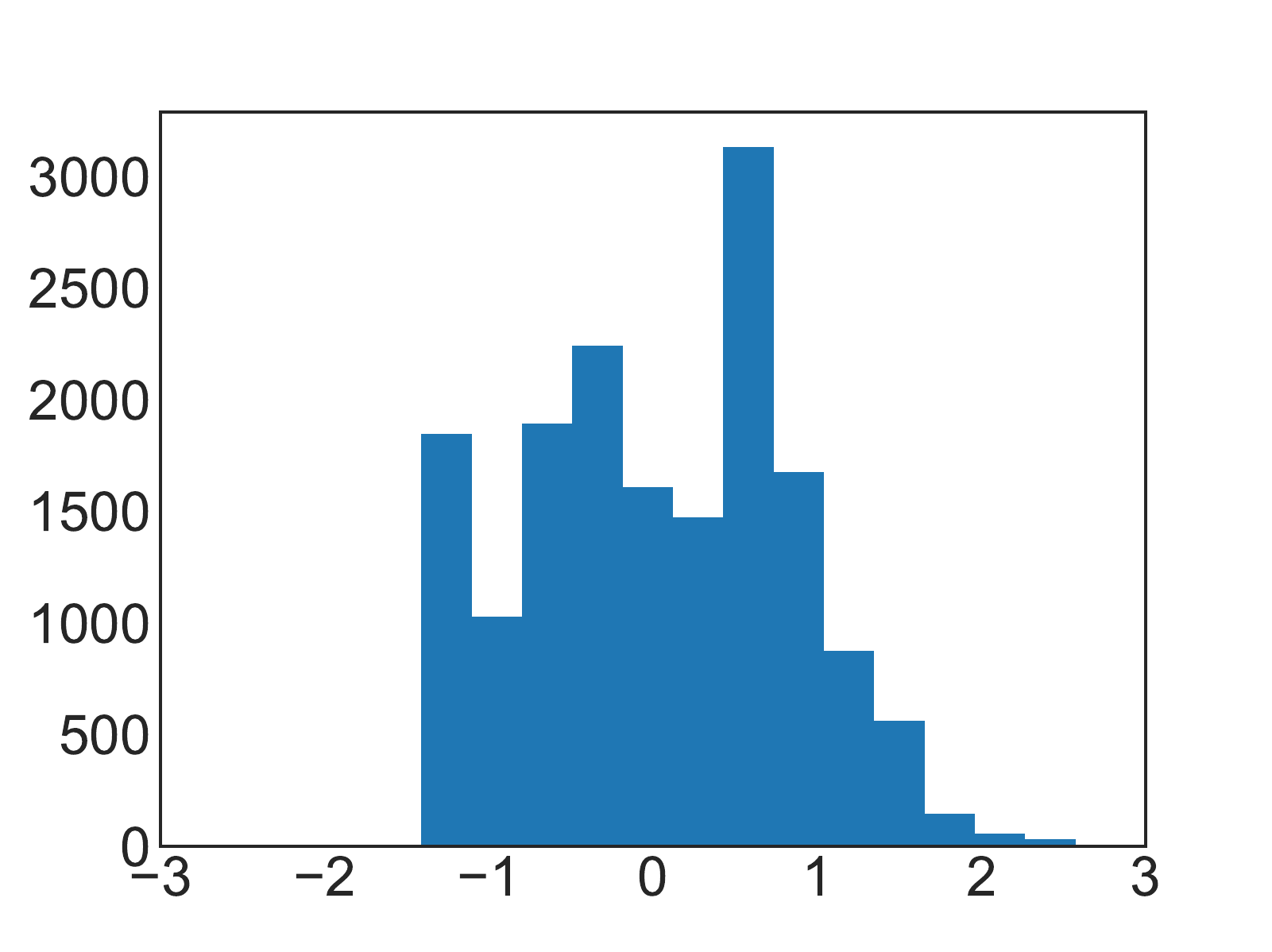}
\vspace{-7.5mm}
\caption{Child's arousal}
\end{subfigure}
\hfill
\begin{subfigure}[h]{0.42\linewidth}
\includegraphics[width=1.00\linewidth]{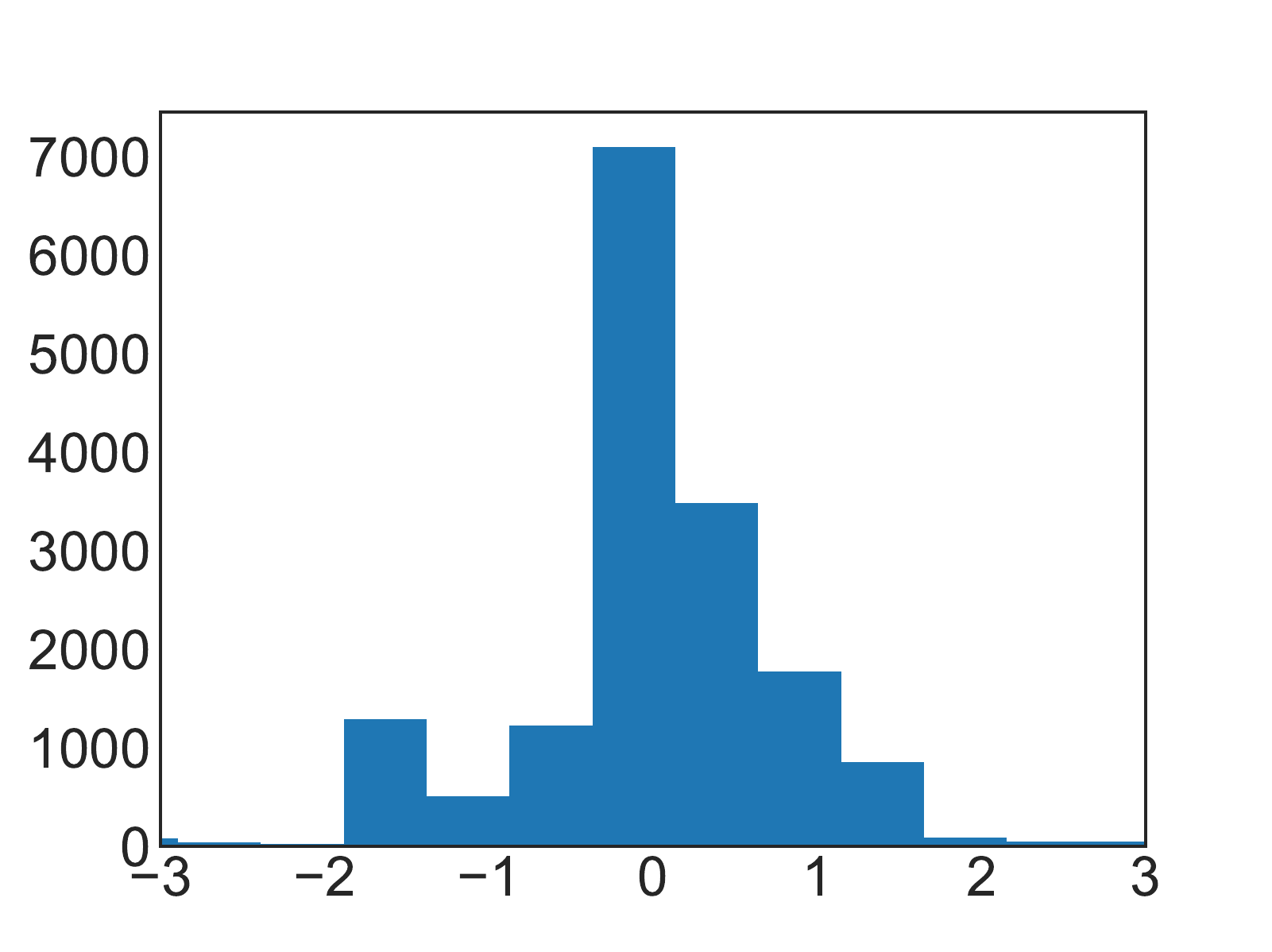}
\vspace{-7.5mm}
\caption{Parent's arousal}
\end{subfigure}
\hfill
\vspace{-0.8mm}
\begin{subfigure}[h]{0.43\linewidth}
 \includegraphics[width=1.0\linewidth]{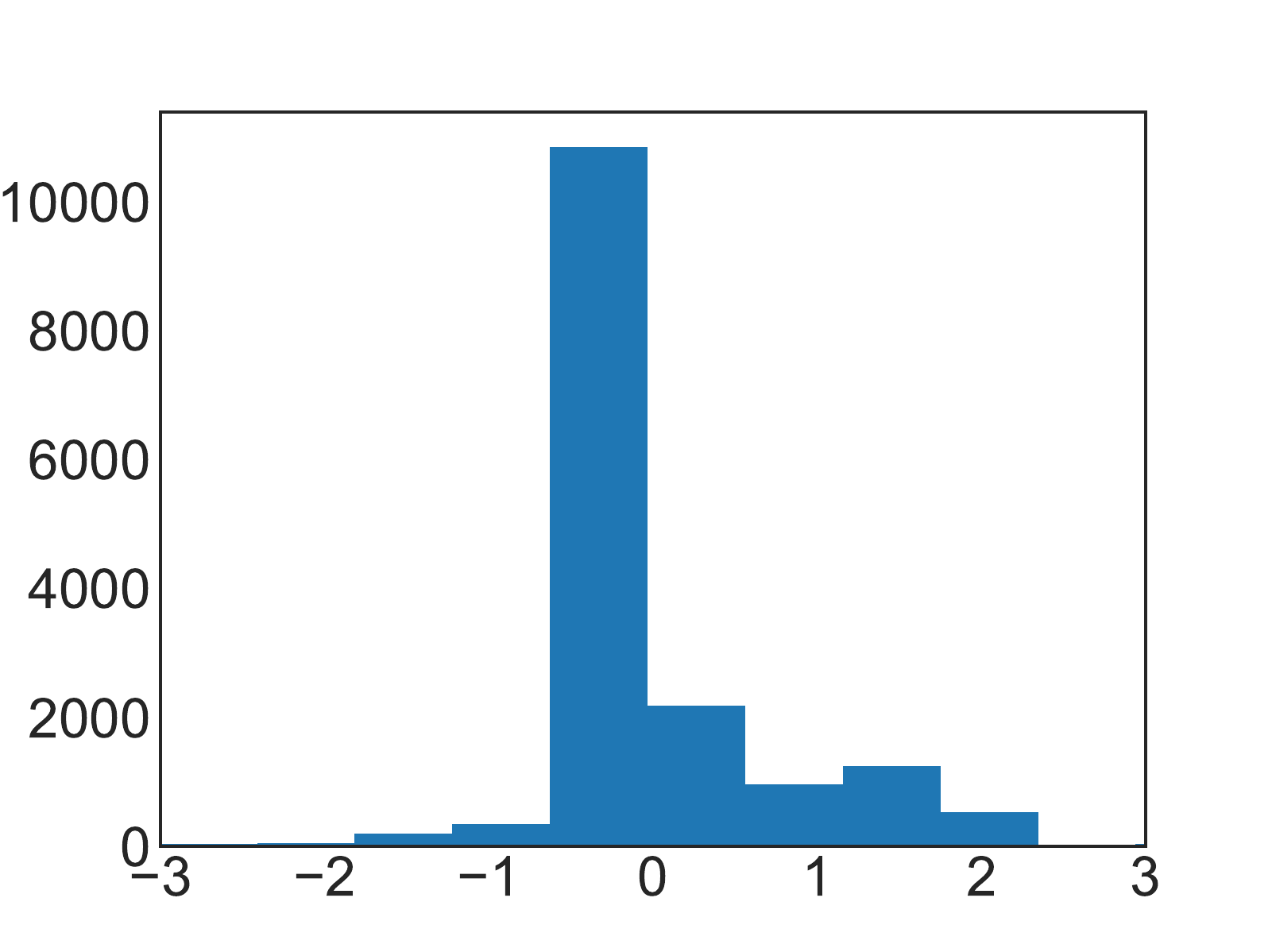}
 \vspace{-7.5mm}
 \caption{Child's valence}

\end{subfigure}
\hfill
\begin{subfigure}[h]{0.42\linewidth}
\includegraphics[width=1.00\linewidth]{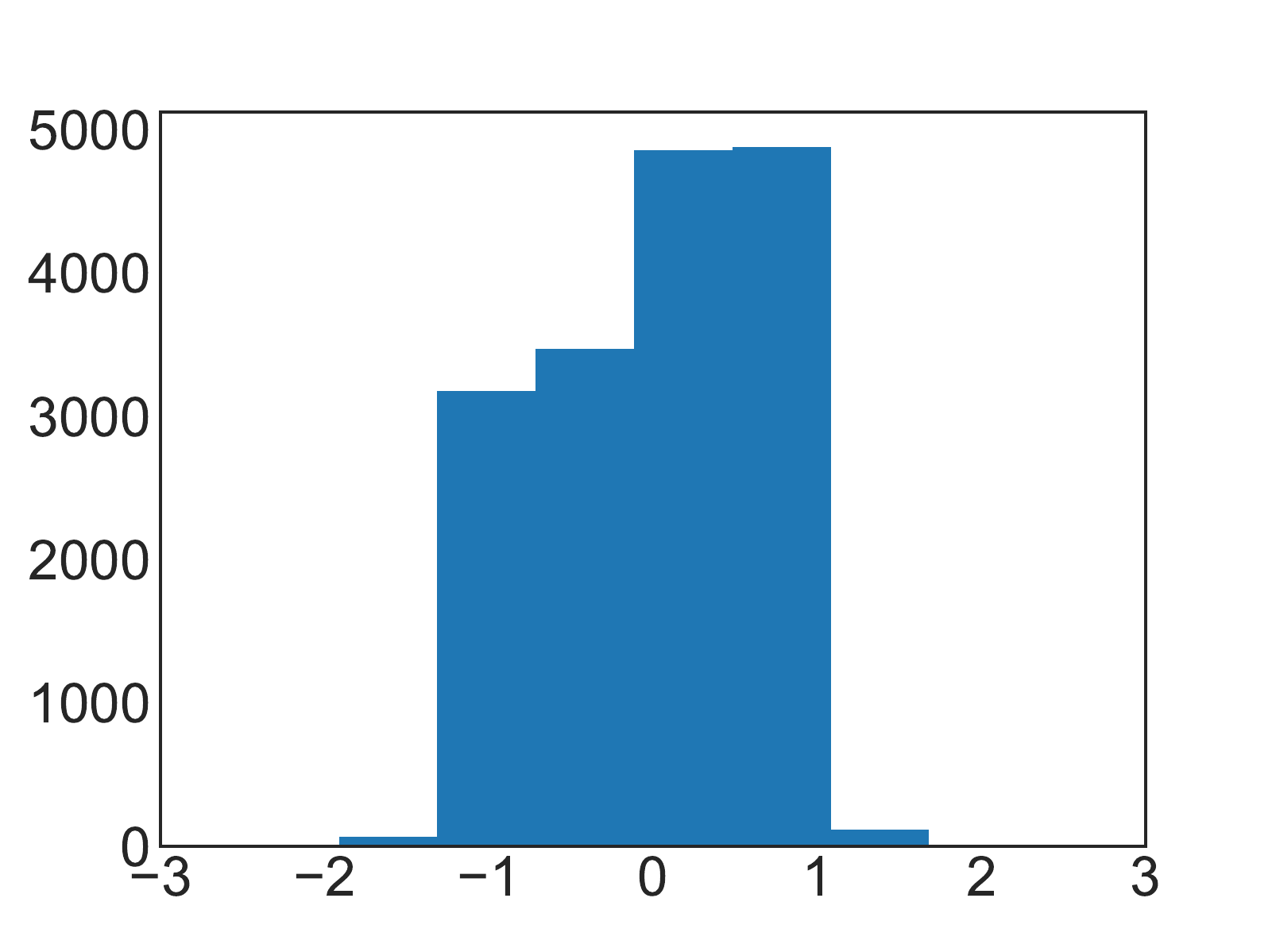}
\vspace{-7.5mm}
\caption{Parent's valence}
\end{subfigure}
\hfill
 \vspace{-4mm} 
\caption{Histograms of the distribution for each affect attribute label after three individual ratings are averaged and standardized.}
\label{fig:histograms-ground-truth-labels}
 \vspace{-5mm} 
\end{figure}

The label distribution for each target attribute after the individual ratings are averaged and standardized is depicted in Figure ~\ref{fig:histograms-ground-truth-labels}, and the pairwise Spearman correlation coefficients between the four affect attributes are reported in Table~\ref{correlation-between-affect-attributes}. The correlation results showed that speaker valence and arousal are moderately positively correlated with each other for both parent and child. Additionally, parent's and child's arousals are negatively correlated with each other. The latter may be explained by the conversational interactions in which the parent and the child took turns being a speaker and listener.

\begin{table}[tbp]
 \caption{Statistics of the number of valence/arousal annotations collected across families. The first, second and third quartiles are denoted as Q1, Q2 and Q3, respectively}
  \label{des-stats-parent-child-dyads}
  \vspace{-3.5mm}
  \begin{small}
  \begin{tabular}{llllll}
  \toprule
   Mean $\pm$ SD & Min & Max & Q1 & Q2 & Q3 \\
   $488.03\pm123.23$ & 235 & 724 & 438.25 & 482.00 & 576.75 \\
  
  \bottomrule
\end{tabular}
 \end{small}
\vspace{-4mm}
\end{table}

\begin{table}[tbp]
 \caption{Left: Pairwise Pearson correlation coefficients between child's arousal (CA), parent's arousal (PA), child's valence (CV), and parent's valence (PV). Right: The average and confidence interval (CI) of intra-class correlation for the four affect attributes.}
  \label{correlation-between-affect-attributes}
  \vspace{-3.5mm}
  \begin{small}
  \begin{tabular}{rrrrr}
  \toprule
   & CA & PA & CV & PV  \\
  \midrule
  CA &1 & & & \\
  PA & -0.27 & 1& & \\
   CV & 0.32 & -0.01& 1 & \\
   PV & -0.19 & 0.43 &0.14 & 1\\
  \bottomrule
\end{tabular}
\quad
 \begin{tabular}{lllll}
  \toprule
   Label  & ICC score & ICC CI  \\
  \midrule
  CA &  0.84 & [0.84, 0.85] \\
  PA &  0.83 & [0.82, 0.83] \\
  CV &  0.64 & [0.63, 0.65] \\
  PV &  0.61 & [0.60, 0.62]\\
  \bottomrule
\end{tabular}
 \end{small}
\vspace{-4.4mm}
\end{table}


\begin{figure*}[t]
\centering
    \includegraphics[width=0.7\textwidth]{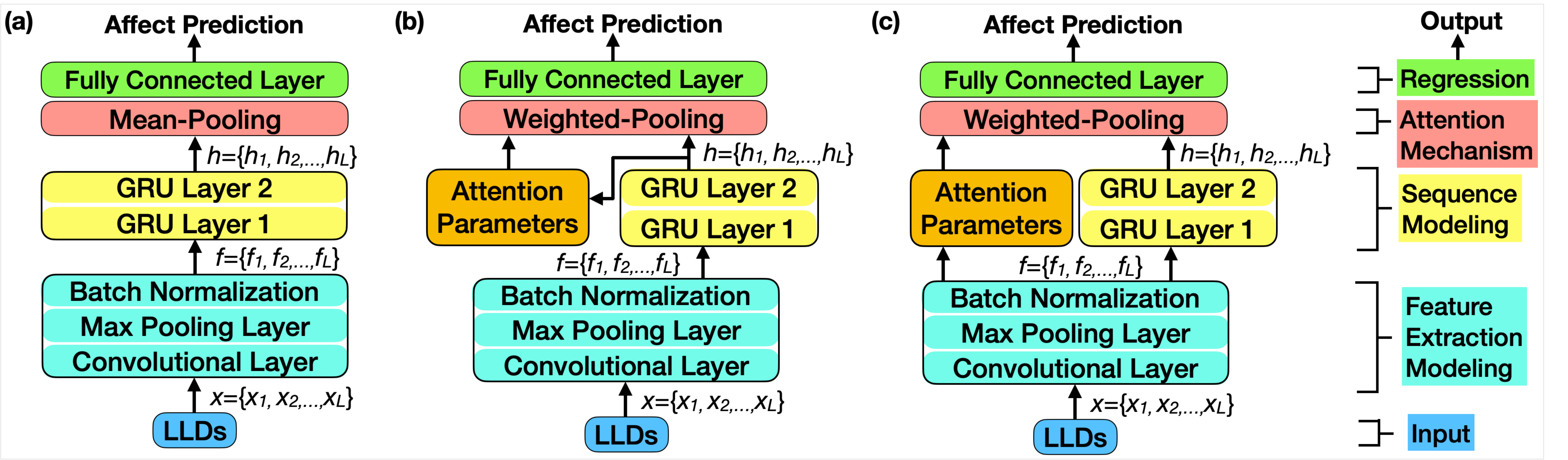}
     \vspace{-4mm}
    \caption{End-to-end convolutional recurrent deep neural networks (CRDNN) architecture design. (a): baseline CRDNN framework without a mean pooling layer (Section~\ref{no-attention-layer}). (b): CRDNN framework with a weighted pooling using the RNN output (Section~\ref{weighted-pooling-hidden-layer}). (c): CRDNN framework with a weighted pooling using the CNN output (Section~\ref{weighted-pooling-acoustic-layer}).}
    \label{fig:end-to-end-learning-architecture}
  \vspace*{-\baselineskip}
\end{figure*}


\section{Methodology}
\subsection{End-to-end speech emotion recognition}
In deep learning, end-to-end architecture refers to the process of learning the direct mapping between the input and final output using a neural network model as a black box and training it according to the global objective function. State-of-the-art end-to-end 
models can be broadly categorized into 1) spectrogram (e.g.,~\cite{Graves2013-SRW}) and 2) raw wave form (e.g.,~\cite{Sainath2015-LTS}) based on the input signal type. Different combinations of neural network models, e.g., convolutional and recurrent, have been explored as end-to-end emotion recognition systems~\cite{Zhao2018-ASO}.

\subsection{Low-level descriptors}
Low-level descriptors (LLDs) are a set of acoustic features computed from short frames of typically 20 to 50 milliseconds. The most relevant feature groups are prosodic, spectral, cepstral, and voice quality features~\footnote{The full list of LLDs computed in this work can be accessed via link (\url{https://drive.google.com/file/d/1Bq6rWLVUggQwyzaJ4c9QU1epOBalCHdr/view?usp=sharing}).}.
The traditional approach to feature extraction is to apply statistical functionals to the LLDs over the duration of the utterance, and then concatenate the aggregated values into a long feature vector at the utterance level to form high-level statistical functions.
Instead, we compute 65 LLDs from each frame of the audio samples using the openSMILE 2.0 toolkit~\cite{Eyben2013-RDI}, and extract temporal information by feeding the frame-level LLDs into the end-to-end learning framework, which systematically crafts the feature set and may save much manual effort. 

\subsection{Model design}
Our end-to-end architecture for parent-child affect recognition consists of convolutional, recurrent, and deep neural networks (CRDNN), as illustrated in Figure~\ref{fig:end-to-end-learning-architecture}. Frame-level LLDs of the raw wave form audios are used as the input signals. As the first module in the architecture, the CNN is used to extract high-level emotional features from LLDs, treating LLDs as two-dimensional images with time and frequency dimensions and reducing the feature engineering. Then, the RNN is used to extract the contextual and time-series information of the CNN output, analyzing the sequence feature of the audio signals. With the spectral and temporal modeling handled by the CNN and RNN layers, an attention layer is used to select regions of speech signals vital for the task of interest. Lastly, the output from the attention layer is fed into a fully-connected (FC) layer, which maps the attention output, i.e., utterance representation, to another space easier for the task of continuous affect recognition.

\paragraph{Input} 
For each five-second audio instance, a set of LLD features is extracted at the frame level, and each LLD is normalized by its global mean and standard deviation calculated across the entire dataset. Then, the input is a sequence of frames $\bm{x}=\{x_1,x_2,...,x_L\}$ where $x_t$ is a set of LLDs at frame $t$ and $L$ is the total number of frames in the five-second audio.

\paragraph{CNN} The CNN structure consists of a single one-dimensional convolutional layer with filters of $1 \times 8$, and takes $\bm{x}$ as the input. Then, it is followed by an overlapping max-pooling layer with pooling size=(1,3) on the frequency axis, and an element-wise rectified linear unit (ReLU). Then, the batch normalization layer~\cite{Ioffe2015-BNA} is applied to the CNN output to speed up training. The final output is denoted as $\bm{f}=\{f_1,f_2,...,f_L\}$.

\paragraph{RNN}
The sequence of higher-level representations from CNNs ($\bm{f}$) is then directed into two bidirectional gated recurrent unit (GRU) layers each having 128 hidden cells to learn a sequential pattern of the input signal data. GRU is chosen for its efficacy of temporal summarization and less demand on training data size in comparison with long short-term memory (LSTM) networks~\cite{Chung2014-EEO}, as well as its effectiveness in modeling individual affect shown in prior work~\cite{Zhang2019}. 
The output sequence of the last recurrent layer is represented as $\bm{h}=\{h_1,h_2,...,h_L\}$. 

\paragraph{Attention layer}
For the models without any attention mechanism, the RNN output is directed into a temporal mean-pooling layer to transform into one single vector $\bm{z}$ for the FC module. For the models with an attention mechanism, an attention layer is applied to transform the RNN output into $\bm{z}$. More details on the attention mechanism are in Section~\ref{attention-mechanism-section}. 

\paragraph{FC}
As a regression module, the FC layer with input size of 128, hidden units of 512, and output size of 1 maps the pooling layer's output, $\bm{z}$, to a final continuous affect score.  

\paragraph{Dropout}
To understand the impact of dropout on model performance, we implement two models with different dropout layer functions for each CRDNN architecture. The first model has one dropout on the CNN layer and one on the RNN. The second model has an additional dropout on the attention layer in addition to the first two dropouts. We do not implement models without CNN and RNN dropout, as prior work suggested applying dropout to the CNN and RNN as a common practice to increase the model’s generalizability~\cite{Srivastava2014-DAS,Gal2015-TGA}. A 10\% dropout is applied to each CNN, RNN, and FC layer during training to prevent overfitting and increase model generalizability. 

\subsection{Attention mechanism}\label{attention-mechanism-section}

We implement the following attention mechanisms. In each architecture, the attention mechanism is added as the layer directly after the second RNN layer and before the FC layer, and the obtained result from the attention mechanism $\bm{z}$ is fed into the FC layer for affect regression.

\subsubsection{Mean-pooling over time without local attention}\label{no-attention-layer}
For the baseline model without local attention, we perform a mean-pooling over time on the RNN outputs ($z$), computed as follows: $$\bm{z}= \frac{\sum_{t=1}^{L} \bm{h_t}}{|L|}$$

\subsubsection{Weighted-pooling with local attention using RNN output}\label{weighted-pooling-hidden-layer}
Instead of the mean pooling over time, we compute a weighted sum of the RNN outputs with the following equation:
$$\bm{z}= \sum_{t=1}^{L}{\alpha_{t}\bm{h}_t}$$ 

\noindent where weight $\alpha_i$ is the weight of the RNN output $\bm{h}_i$. The equation to compute $\alpha_{t}$ is as follows:     

$$\alpha_t= \frac{exp(\bm{w}^T\bm{h}_t)}{\sum_{i=1}^{L}{exp(\bm{w}^T\bm{h}_i)}}$$

At each time frame $t$, the inner product between the learnable parameter $\bm{w}$ and the RNN output $\bm{h_t}$ is interpreted as a score for the contribution of frame $t$ to the final utterance-level resentation of the affect. Then, the result is fed into a softmax function to obtain a set of final weights $\alpha_{t}$ which sum to unity. 

\subsubsection{Weighted-pooling with local attention using the CNN output}\label{weighted-pooling-acoustic-layer}
Instead of learning to weight the frames based on the signal's temporal dynamic from the RNN output, we design a new weighted-pooling layer that allows the model to weight the frames based on speaker characteristics, which are mostly present in the input features or lower layers of the network. This weighted pooling with the CNN output is similar to the one with the RNN output except that the weight of the RNN output $\alpha_t$ is calculated using the CNN output ($\bm{f}_t$) instead of the RNN output ($\bm{h}_t$), as follows: 



$$\alpha_t= \frac{exp(\bm{w}^T\bm{f}_t)}{\sum_{i=1}^{L}{exp(\bm{w}^T\bm{f}_i)}}$$

\section{EXPERIMENTS}
In this study, we focused on the audio modality in the collected audio-visual dataset described in Section~\ref{data collection}. The dataset has four affect attribute categories: parent's arousal and valence, child's arousal and valence. All four categories were used to train the deep models. 

\subsection{Evaluation settings}
To evaluate the different deep learning architectures, we used a leave one speaker group out cross validation (LOSGO-CV). We partitioned 34 dyads into five non-overlapping groups. The evaluation setting was subject-independent; therefore, data instances (five-second audio fragments) from each dyad only appeared in one of the five groups. When partitioning the dataset, each group's size was additionally balanced by sorting dyads based on their instance sizes and assigning the largest dyad in the unassigned dyad list to each group iteratively. For each iteration, we selected one as the development (dev) set, one as the test set and the rest as the train set. The experiment repeated until each group had been used as dev and test sets once, and the experiment was repeated five times in total. In total, each experiment was trained on a training set of $9955.8\pm327$ instances, a dev set of $3318.6\pm255$ instances and a test set of $3318.6\pm255$ instances. We plan to publish the split with the dataset for other researchers to use the same split for model comparisons. To evaluate model performance, the Spearman's rank correlation coefficient ($\rho$) was selected for its ability to assess monotonic relationships (whether linear or not) without assuming normal distributions of variables, given that the four affect annotations did not have normal distributions. Note that $\rho$ ranges from [-1,1]; however, we report them in \%, i.e., [-100,100].

\subsection{CRDNN Model training}
Our CRDNN models were implemented using TensorFlow, and different models were trained separately for each speaker and each affect attribute. The training of the proposed architectures was conducted using the Adam optimization algorithm with an initial learning rate of 0.0001 and a batch size of 20. For the loss function, we used the concordance correlation coefficient ($ccc$), which measures agreement as a departure from perfect linearity (y = \^{y}). 
An early stop was implemented to avoid overfitting issues. Specifically, the training would stop if the model's performance on the dev dataset did not improve after 15 epochs.

\begin{table}[tbhp]
 \caption{Model performance on predicting child's arousal}
  \label{child-arousal-result}
  \vspace{-3.5mm}
  \begin{small}
  \begin{tabular}{lllll}
  \toprule
  Methods  & \multicolumn{1}{c}{DROP(ALL)} & \multicolumn{1}{c}{DROP(CR)}  \\
       & $\rho$ ($mean\pm SD$)  & $\rho$ ($mean\pm SD$)  \\
  \midrule
  ATT(NO)     & $52.1\pm6.6\%$ & $53.0\pm7.2\%$ \\
    \midrule
ATT(R)    & $53.5\pm6.8\%$ &  $\bm{54.7\pm5.9\%}$**$^\dagger$ \\
  \midrule
  ATT(C)    & $\bm{54.0\pm7.8\%}$*  & $51.9\pm9.0\%$ \\

  \bottomrule
\end{tabular}
 \end{small}
\vspace{-7mm}
\end{table}

\begin{table}[tbhp]
 \caption{Model performance on predicting parent's arousal.} 
  \label{parent-arousal-result}
  \vspace{-3.5mm}
  \begin{small}
  \begin{tabular}{lllll}
  \toprule
   Methods  & \multicolumn{1}{c}{DROP(ALL)} & \multicolumn{1}{c}{DROP(CR)}  \\
       & $\rho$ ($mean\pm SD$)  & $\rho$ ($mean\pm SD$)  \\
  \midrule
  ATT(NO)    &   $53.6\pm9.8\%$ &  $53.5\pm9.9\%$ \\
    \midrule
ATT(R)     & $\bm{53.7\pm11.7\%}$ &  $\bm{53.8\pm10.0\%}$\\
  \midrule
  ATT(C)  & $53.1\pm9.9\%$&  $53.6\pm9.6\%$ \\

  \bottomrule
\end{tabular}
 \end{small}
\vspace{-7mm}
\end{table}

\begin{table}[tbhp]
 \caption{Model performance on predicting child's valence.} 
  \label{child-valence-result}
  \vspace{-3.5mm}
  \begin{small}
  \begin{tabular}{lllll}
  \toprule
   Methods  & \multicolumn{1}{c}{DROP(ALL)} & \multicolumn{1}{c}{DROP(CR)}  \\
       & $\rho$ ($mean\pm SD$)  & $\rho$ ($mean\pm SD$)  \\
  \midrule
  ATT(NO)      & $\bm{22.1\pm6.4\%}$* &  $21.4\pm4.2\%$ \\
    \midrule
ATT(R)    & $19.8\pm5.2\%$  & $19.5\pm5.8\%$\\
  \midrule
  ATT(C)    & $20.4\pm5.9\%$ &  $\bm{21.7\pm4.2\%}$ \\

  \bottomrule
\end{tabular}
 \end{small}
\vspace{-7mm}
\end{table}

\begin{table}[tbhp]
 \caption{Model performance on predicting parent's valence.} 
  \label{parent-valence-result}
  \vspace{-3.5mm}
  \begin{small}
  \begin{tabular}{lllll}
  \toprule
  Methods  & \multicolumn{1}{c}{DROP(ALL)} & \multicolumn{1}{c}{DROP(CR)}  \\
       & $\rho$ ($mean\pm SD$)  & $\rho$ ($mean\pm SD$)  \\
  \midrule
  ATT(NO)      & $\bm{28.1\pm16.1\%}$* &   $31.6\pm14.4\%$ \\
    \midrule
ATT(R)     & $26.1\pm17.3\%$ &  $30.4\pm14.3\%$\\
  \midrule
  ATT(C)    & $26.7\pm14.6\%$ &  $\bm{33.7\pm12.4\%}$**$^\dagger$\\

  \bottomrule
\end{tabular}
 \end{small}
\vspace{-4.5mm}
\end{table}

\section{Results and Discussion}
To compare the performance of different CRDNN architectures, we trained and evaluated 24 models in total with six models per learning task. Models are categorized into three attention types: (1) ATT(NO) denotes no attention layer models with a mean-pooling layer (Section~\ref{no-attention-layer}), (2) ATT(R) denotes models with a weighted-pooling layer using the RNN output (Section~\ref{weighted-pooling-hidden-layer}), and (3) ATT(C) denotes models with a weighted-pooling layer using the CNN output (Section~\ref{weighted-pooling-acoustic-layer}). For each attention implementations, we trained two models, which are the DROP(ALL) model that has dropouts on its CNN, RNN and attention layers, and the DROP(CR) model that has dropouts on its CNN and RNN layers but not on its attention layer. 
Results showing Spearman correlation coefficient means and standard deviations obtained on the six models for child's arousal, parent's arousal, child's valence, and parent's valence are reported in Tables~\ref{child-arousal-result}, ~\ref{parent-arousal-result}, ~\ref{child-valence-result} and ~\ref{parent-valence-result}, respectively. 

When comparing the performance between two models, we used a z-test to evaluate the statistical significance of the difference in their Spearman correlation coefficients ($\rho$). The sample size in the z-test was the total dataset size ($N=16,593$), as every data instance was used as the test instance once in the cross-validation \cite{Dietterich1998-AST}. In each learning task, we evaluated the performance difference between the ATT(NO) model and the most effective attention-augmented model within the same dropout configuration, and the z-test results with $p<0.05$ significance level marked with $*$ and $p<0.001$ as $**$ in the tables. Additionally, we evaluated the performance difference between the most effective ATT(NO) model and attention-augmented model across both DROP(ALL) and DROP(CR) configurations in each task, and the overall most effective models with the significant z-test results are marked with $^\dagger$ in the tables. 


\begin{figure}[th]
\centering
\begin{subfigure}[h]{0.496\linewidth}
 \centering
 \includegraphics[width=1.0\linewidth]{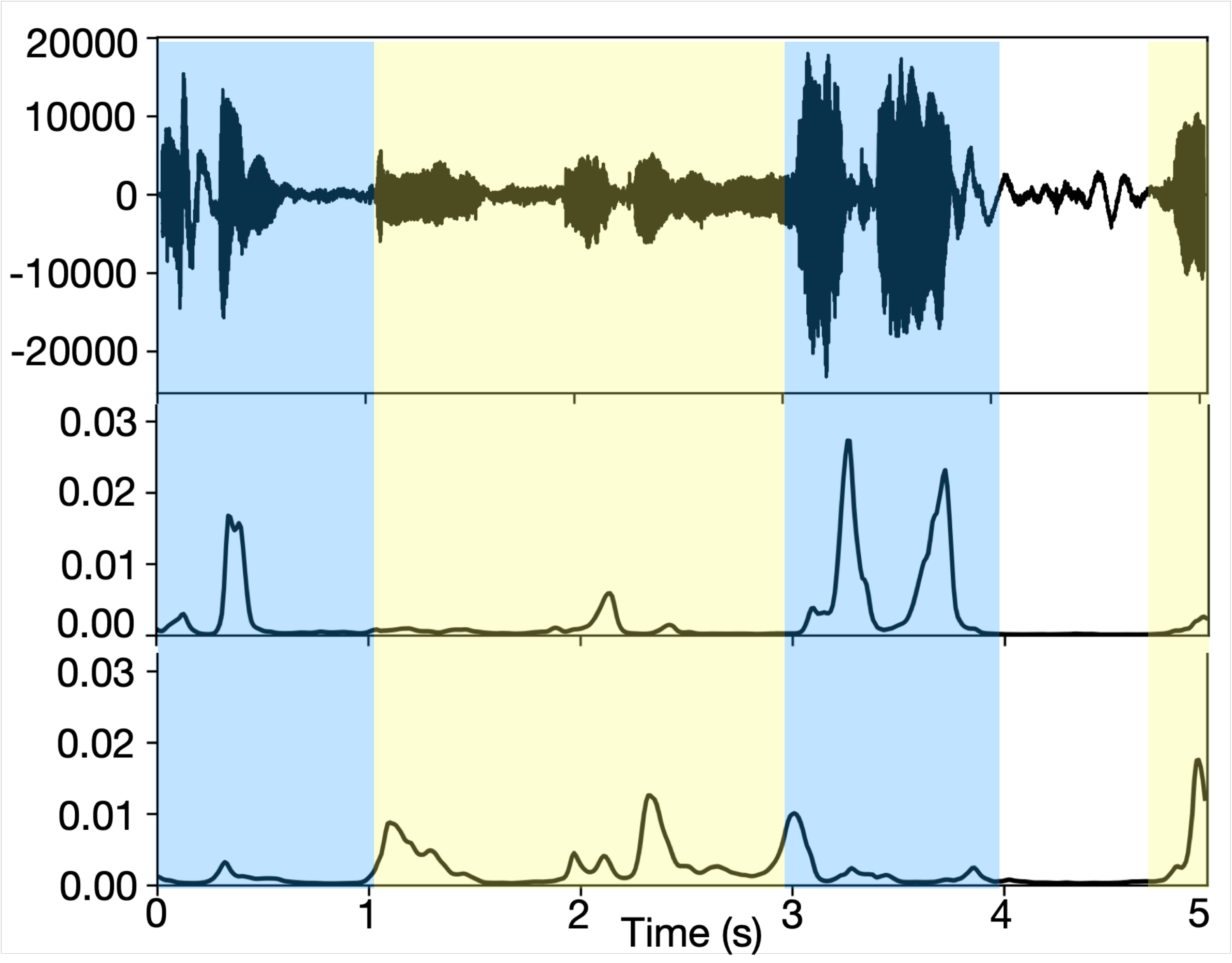}
 \label{fig:attention-weights-child-arousal}
\end{subfigure}
\begin{subfigure}[h]{0.4936\linewidth}
\centering
\includegraphics[width=1.0\linewidth]{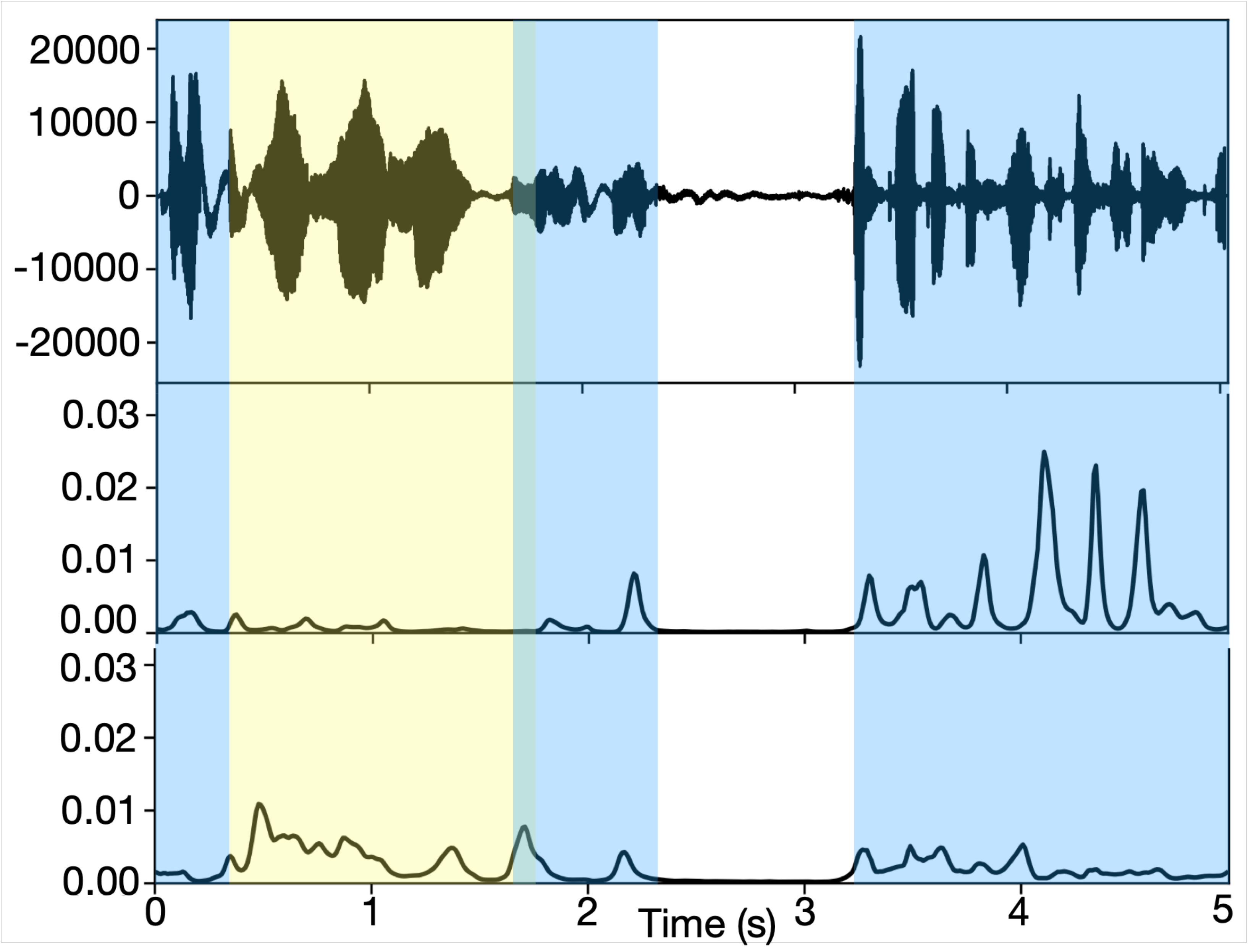}
\label{fig:attention-weights-parent-arousal}
\end{subfigure}
 
 \vspace{-8mm}
\caption{The attention-augmented model, ATT(R)+DROP(CR), automatically learned attention weights for child's and parent's arousal predictions; results are shown here for two audio wave files from the test set. The time region for parent's and child's utterances are highlighted in blue and yellow, respectively. The top, middle and bottom subplot are the amplitude of the raw audio signals and attention weights for parent's arousal model, and weights for child's arousal model, respectively.}
\label{fig:attention-weights}
  \vspace{-6.5mm}
\end{figure}

\subsection{Model performance across learning tasks}
\subsubsection{Comparisons between arousal and valence} All six models achieved significantly higher $\rho$ scores on child's arousal and parent's arousal than on child's valence and parent's valence, according to the z-test results ($p<0.05$). The most effective models for child's arousal and parent's arousal achieved $54.7\pm5.9\%$ and $53.8\pm10.0\%$, respectively; conversely, the most effective models for valence achieved only $22.1\pm6.4\%$ and $33.7\pm12.4\%$. This performance difference between predicting a speaker's arousal and valence is consistent with findings from prior studies in affect recognition research~\cite{Trigeorgis2016-AFE,Zhang2019-AAE}, which have repeatedly shown that valence is much more difficult to predict from speech compared to arousal. 

\subsubsection{Comparisons between parent's and child's valence} All six models achieved significantly higher $\rho$ scores when predicting parent's valence than predicting child's valence according to the z-test results ($p<0.05$). The $\rho$ scores across the six models are $29.4\pm3.0\%$ and $20.8\pm1.1\%$ on average for parent's and child's valence, respectively. This finding suggests that it was easier to predict parent's valence than child's valence, and such difference could be explained by how parents and children interacted when reading stories together. Based on our observation, parents in our study were more verbally expressive and made more utterances than children did, making it easier for models to pick up their valence from their verbal behaviors. Additionally, children in our study were diverse in their personalities and temperaments, and several of them tended to exhibit their valence predominantly through body gesture and facial expression without making utterances, particularly when they were listening to their parents reading stories. This would make it more difficult for a single-modality speech model to predict children's valence. The finding creates a higher motivation to integrate children's developmental profiles (e.g., temperament) as an additional modality into future dyadic affect model design. 

   
\subsection{Effect of attention mechanism on model performance}
\subsubsection{Overall performances of attention-augmented models}
In child's arousal and parent's valence tasks, the most effective attention-augmented model achieved significantly higher $\rho$ scores than the most effective attention-free model, as analyzed by the z-test (child's arousal: $p<0.001$; parent's valence: $p<0.001$). In the other two tasks, the most effective attention-augmented and attention-free models achieved comparable performances with their z-test results not being statistically significant (parent's arousal: $p>0.05$; child's valence: $p>0.05$). This result showed that integrating a weighted attention layer into a CRDNN model led to its improved performance on predicting child's arousal and parent's valence in parent-child interactions, indicating the overall effectiveness of using an attention mechanism in dyadic affect recognition despite that statistically significant performance gains may not be evident in all affect learning tasks.

\subsubsection{Parent's and child's arousal}
A comparison of model performance between parent's arousal and child's arousal tasks showed that integrating a weighted attention layer into a CRDNN model led to a larger performance improvement on child's arousal than parent's arousal as seen in Tables~\ref{child-arousal-result} and ~\ref{parent-arousal-result}. For child's arousal, the most effective attention-augmented model  ATT(R)+DROP(CR) achieved $54.7\%$, showing a statically significant increase of 1.7\% over the most effective attention-free model ($\rho=53.0\%$), according to the z-test ($p<0.001$). In addition, the most effective attention-augmented models in the two dropout configurations, that is, ATT(C)+DROP(ALL) and ATT(R)+DROP(CR),  significantly outperformed their attention-free counterparts in child's arousal task ($p<0.05$), respectively. This finding further strengthens the claim that having a weighted-pooling attention layer would improve a CRDNN model's prediction on child's arousal. Conversely, in the parent's arousal task, the performance difference between the most effective attention-augmented and attention-free models was not statistically significant in each dropout configuration. 

To further investigate how the weighted-pooling attention layer in the proposed framework was trained, we illustrate the learned attention weights of ATT(R)+DROP(CR) model in parent's and child's arousal tasks for two wave audio files from the test set along with the audio files' speech amplitudes, seen in Figure~\ref{fig:attention-weights}. Generally speaking, as shown in the weight distributions of both child's and parent's arousal models in each audio, one speaker's high amplitude was not picked up by the other speaker's model and vice versa. Additionally, the model identified the correct speaker, i.e., implicitly learned speaker diarization, with a speaker's weight scoring higher in the parts of signals having a high amplitude of the target speaker's utterance, a signal of speaker's arousal.   

The weighted-pooling layer's ability to identify regions of signals relevant to individual affect led to a statistically significant improved performance on predicting child's arousal but not parent's arousal. It could be partially explained by the dyadic interaction styles in our study. In most dyads in our dataset, the parent made more utterances than the child did because the parent read stories and led the story-related discussions. As a result, it became comparatively less crucial for models to accurately pick up the specific parts of signals relevant to parent's utterances, as averaging entire audio signals could sufficiently reflect the parent's arousal. However, each five-second instance might have only short time regions of signals relevant to a child's arousal, with the rest containing noisy frames (e.g., parent's utterances). In the child's arousal task, identifying the correct time regions and filtering out the noisy regions thus became more crucial for prediction, and the importance of the weighted-pooling layer became more evident. From the model design perspective, it was possible that the RNN layer already learned to focus on the correct speaker, particularly when one speaker made more utterances than the other in an audio; therefore, the hidden state from the RNN layer used to train the attention layer did not contain much information from the speaker who made fewer utterances, and averaging the hidden states was sufficient for predicting the arousal of the dominant speaker, i.e., parent's arousal.

\subsubsection{Parent's and child's valence}
For parent's valence, the most effective model, ATT(C)+DROP(CR), achieved a $\rho$ value of $33.7\%$, outperforming the most effective attention-free model ($\rho=31.6\%$) by 2.7\% with a statistically significant z-test result ($p<0.05$). For child's valence, the performance difference between the most effective attention-augmented and attention-free models was not significant. 
We do not show learned attention weights associated with raw speech amplitudes for valence given that speech amplitude is known to have little association with valence (unlike arousal, which often changes with speech amplitude~\cite{Eyben2015-RSA}). 

In both valence tasks, the weighted-pooling layer using the CNN output, ATT(C), yielded significantly higher $\rho$ scores than the weighted-pooling layer using the RNN output,  ATT(R). This finding was, however, not evident in the two arousal tasks, and evidenced that lower-level speaker's speech characteristics directly extracted from the CNN layer contained information more crucial for the weighted pooling layer to pick up time regions vital to speaker valence than the temporal domain context information processed by the RNN layer. Overall, this finding suggests that, when predicting individual affect in dyads, it is important not to assume that an attention mechanism could work equally effectively for both valence and arousal tasks and the effectiveness of an attention mechanisms depends on the affect prediction task (either valence or arousal).

\subsection{Effect of dropout on model performance}
In four learning tasks, the performance difference between DROP(ALL) and DROP(CR) models was not consistently observed. In a few cases, DROP(CR) models outperformed DROP(ALL) models. For example, the performance of ATT(R) model on child's arousal improved from 53.5\% to 54.7\% when using DROP(CR) instead of DROP(ALL), a difference that is statistically significant ($p<0.05$). A similar trend was observed for parent's valence where ATT(C)+DROP(CR) and ATT(R)+DROP(CR) achieved 33.7\% and 30.4\%, respectively, while with DROP(ALL) only  26.7\% and 26.1\% was achieved. These two differences are also both statistically significant ($p<0.001$). Only in the case of using the ATT(C) model for child's arousal, having a dropout function on the model's attention layer significantly improved its performance with an increase from 51.9\% to 54.0\% ($p<0.001$). This finding suggests that the dropout function on a CRDNN model's attention layer does not necessarily improve its performance, and may even impede its prediction quality in some affect recognition tasks.  

In three of four learning tasks, DROP(CR) models had smaller variability in the correlation scores ($\rho$) during a five-fold cross validation compared to DROP(ALL) models. Specifically, the average standard deviation scores of $\rho$ values dropped from 5.8\% to 4.7\%, 16.0\% to 13.7\%, 10.4\% to 9.8\% when the dropout function on the attention layer was removed from models prediction of child's valence, parent's valence, parent's arousal, respectively. This effect was more evident in the attention-augmented models than in the attention-free models. 
This finding suggests that having a dropout function on the attention layer could potentially reduce the robustness of a CRDNN model and lead to greater variability in its performance on unseen datasets. Overall, a dropout function on the attention layer is not recommended for future model design, particularly when the attention layer is implemented using weighted pooling.

\section{Conclusion}
Results on the new DAMI-P2C dataset show that an end-to-end learning framework can achieve competitive performance in predicting the affect of individuals in a dyadic interaction. With a weighted pooling layer added between the RNN and FC layers as the model's attention layer, the end-to-end system significantly improved its performance in terms of the Spearman correlation coefficient on predicting child and parent arousal levels. The most effective attention mechanism was different across the four learning tasks:  The weighted-pooling layer using the RNN output led to a larger performance increase for child's arousal, whereas the weighted-pooling layer using the CNN output was more effective for parent's valence. Lastly, we found that the attention-augmented models are able to focus on the target speaker, providing implicit diarization of who speaks when.  

We are open-sourcing this dataset of parent-child dyadic co-reading interaction, which contains LLDs of the 34 dyads' audios, arousal and valence annotations, and a diverse set of developmental, social and demographic profiles of each dyad. 
Participant families came from diverse social and cultural backgrounds, allowing for the exploration of building affect sensing systems personalized to individual variations in affect across dyads. The dataset is a unique contribution to the study of modeling multi-speaker speech affect in human-human interactions, and we hope that the end-to-end deep learning systems designed in this work provide a competitive affect recognition baseline for future research in this field. Lastly, we also plan to open-source the video and tablet-based interaction log data in future, and this multimodal dataset would further increase the applicability of affect detectors to 
challenging real-world cases (e.g., handling technical difficulties such as background noise and poor lighting)~\cite{BOSCH2015AAH} as well as recognize a greater variety of affective states (e.g., mind wandering~\cite{STEWART2017FFD,STEWART2016WYM}).  

\section{Limitations and Future work}
This work has several limitations. Since we did not directly compare the attention-augmented models trained on speaker-blind data with models trained on speaker-diarized data, our experimental results do not quantify how accurately the models are able to diarize speakers. Instead, our findings suggest only that the attention-augmented models' higher performances on predicting child's arousal and parent's valence could be attributed to their attention mechanisms that implicitly diarize speakers when picking up parts of signals vital for the learning task of interest. This indication provides an increased motivation for such direct comparisons between speaker-blind and speaker-diarized models in future work. Secondly, our system takes frame-level acoustic features as its input, but it can be extended to directly take the raw waveform by adding a time-convolutional layer~\cite{Sainath2015-LTS}, which can learn an acoustic model and transform the raw time-domain waveform to a frame-level feature. This extension would further reduce our system's reliance on offline pre-processing and enlarge an intelligent agent's capability of sensing and reacting to human affect in real-time interactions. 

Our dataset provides many opportunities to analyze and model multi-speaker affect in parent-child co-reading interactions. For example, an end-to-end neural speaker diarization~\cite{Fujita2019-ETE} or multi-task learning~\cite{Zhang2019-AAE} can be integrated into our current affect prediction systems to jointly learn speaker diarization and individual speaker's valence and arousal. 
Further, the demographic and developmental profiles can be integrated into the current system to model multi-speaker affect in a personalized or culture-sensitive manner ~\cite{Rudovic2018-CDL,Rudovic2018-PML}, as individual differences modulate affective expressions suggested by a theoretical foundation for affect detection~\cite{DMello2018MMA} and cultural differences were empirically found to exist in some aspects of emotions, particularly emotional arousal level~\cite{Lim2016CDE}. 
Computational models that integrated a child's cultural background as an additional input modality were also found more effective in modeling children's affect and engagement than models that did not~\cite{Rudovic2018-CDL,Rudovic2018-PML}. Thus, personalized or culture-sensitive affect models may further improve model prediction performance, and help this technology to faciliate better parent-child dyadic experiences in the future.

\section{Acknowledgement}
This work was supported by the MIT Integrated Learning Initiative (MITili) grant and partly by the Intel Graduate Fellowship.

\bibliographystyle{ACM-Reference-Format}
\balance
\bibliography{bib_bk.bib}

\end{document}